%
%                               April 19, 1999.
%
%                               J.P. Rodriguez
%                               Dept. of Physics and Astronomy
%                               California State University at L.A.
%                               Los Angeles, CA 90032.
%
%                               Tel.: 323-343-2133
%                               Fax.: 323-343-2497
%                               Internet: jrodrig@calstatela.edu
%
%
% Editors
% Physical Review B
% P.O. Box 1000,
% Ridge, New York 11961.
%
%
% Dear Editors,
%
% Please find enclosed an erratum to my article entitled
% ``Thermal properties of gauge fields common to anyon 
% superconductors and spin liquids'', which is published in 
% Phys. Rev. B, vol. 51, pg. 9348 (1995).  The erratum is
% composed in PLAIN TeX.  I thank you in advance for 
% considering my submission.
%
% Sincerely your, J.P. Rodriguez.
%
%
%%%%%%%%%%%%%%%%%%%%%%%%%%%%%%%%%%%%%%%%%%%%%%%%%%%%%%%%%%%%%%%%%%%%%%%%%%%%%
%
% TEX FILE: ERRAT3.TEX (April 19, 1999)
\magnification=1200
\baselineskip=17pt

\centerline {\bf Erratum:   ``Thermal properties of gauge fields common}
\centerline 
{\bf \quad\qquad\qquad to anyon superconductors and spin liquids'',} 
\centerline
{\bf \quad\qquad[Phys. Rev. B {\bf 51}, 9348 (1995)]}
\centerline
{\it \quad\qquad  J.P. Rodriguez}
\bigskip\bigskip\bigskip
The estimate $ k_B T_c\sim g_0\hbar\omega_0$ given in the paper
for the deconfinement temperature
of compact QED$_3$ in the weak-coupling limit, 
$g_0\rightarrow 0$, is wrong.
The factor of $\beta^{N_m}$ that appears in the
two-dimensional (2D)
Coulomb gas (CG) ensemble (Eq. 4)
must   be replaced by $(\beta y_0)^{N_m}$,
where $y_0 = {\rm exp}(-{\rm const.}/g_0^2)$ is the bare
fugacity of a single instanton (see refs. 20 and 21).  
Since the weak-coupling regime,  $g_0\ll 1$,  then
corresponds to an exponentially small 
effective fugacity $y = \beta y_0$ of  the 2D CG (Eq. 4), we
obtain a deconfinement transition at
$2\pi/\beta g_0^2 = 4$.
%that is dual to that of the 2D CG.
The correct value,
$$k_B T_c = {2\over \pi} g_0^2 \hbar\omega_0,$$
for the critical temperature
is therefore notably {\it less}
than that of the strong-coupling limit,
$k_B T_c = 0.73 g_0^2 \hbar\omega_0$, 
which confirms  the 
general expectation (see ref. 15).
The above correction does not affect any of the
conclusions drawn in the paper.

\end